\documentclass[aps,prb,twocolumn,floatfix,showpacs]{revtex4}
\usepackage{color}
\usepackage{epsfig}
\usepackage{graphicx}
\usepackage{amssymb}
\usepackage{amsmath}

\def\ep{{\epsilon}}
\def\cg{{\cal G}}

\def\om{{\omega}}

\def\R{{{\bf R}}}

\def\nnu{{\nonumber}}

\definecolor{darkred}{rgb}{0.75,0.0,0.0}
\def\g{{\bf{g}}}

\def\beq{\begin{equation}}
\def\eeq{\end{equation}}
\def\beqa{\begin{eqnarray}}
\def\eeqa{\end{eqnarray}}

\def\g0{{\gamma_0}}

\def\dna{{\downarrow}}
\def\upa{{\uparrow}}

\newcommand\bear{\begin{eqnarray}}
\newcommand\eear{\end{eqnarray}}
\newcommand\bea{\begin{align}}
\newcommand\ena{\end{align}}
\begin{document}
\title{Field dependent dynamics in the metallic regime of the 
half-filled Hubbard model.}
\author{D.\ Parihari*, N.\ S.\ Vidhyadhiraja** and A.\ Taraphder*$\dagger$}
\affiliation{Centre for Theoretical Studies* and Department of Physics*$\dagger$\\ 
Indian Institute of Technology, Kharagpur 721302, India
\\Theoretical Sciences Unit,\\Jawaharlal Nehru Centre For
Advanced Scientific Research,\\Bangalore 560064, India**}

\date{\today}
\widetext

\begin{abstract}
{ 
A systematic study of the effect of magnetic field (h) on Hubbard model
has been carried out at half filling within dynamical mean field theory.
In agreement with previous studies, we find a zero temperature 
itinerant metamagnetic transition, reflected in the discontinuous changes in magnetization as well as in the hysteresis,
from a paramagnetic (PM) metallic state to a 
polarized quasi-ferromagnetic (QFM) state,
at intermediate and large interaction strength ($U$). The jump in 
magnetization
vanishes smoothly with decreasing interaction strength, and at a critical 
$U$, the transition becomes continuous. The region of `coexistence' of the
PM and QFM solutions in the field-$U$ plane obtained in this study 
agrees quantitatively with recent numerical renormalization group calculations, 
thus providing an
important benchmark. We highlight the changes in 
dynamics and quasiparticle weight across this transition.
The effective mass increases sharply as the transition is approached, 
exhibiting a cusp-like singularity at the critical field,
and decreases with field monotonically beyond the transition.
We conjecture that the first order metamagnetic transition is a result
of the competition between Kondo screening, that tries to quench the
local moments,  and Zeeman coupling, which induces polarization and hence
promotes local moment formation.
A comparison of our theoretical results with experiments on 
$^{3}He$ indicate that, a theory of $^{3}He$ based on the half-filled
Hubbard model places it in a regime of intermediate interaction strength. 
}
\end{abstract}
\maketitle

\section{Introduction}

One of the most interesting aspects of strongly correlated electrons is
the self-consistent emergence of a low energy scale~\cite{hewson}. This 
has been 
demonstrated over the years from a large number of theoretical and
experimental studies. The Kondo quenching of local moments and the 
emergence of a Fermi liquid below a critical temperature in Mott-Hubbard
systems is clearly seen in dynamical mean-field theory (DMFT). A question, 
therefore, naturally arises as to how an external magnetic field would 
interact with the local moments and how the Mott-Hubbard physics gets 
affected by this added field that couples with the moments. 

The Mott-Hubbard systems like $(V_{1-x}Cr_{x})_{2}O_{3}$~\cite{menth}, 
$V_{2}O_{3}$~\cite{sundar} display local magnetic moments which also 
interact through residual antiferromagnetic exchange. There is 
a competition between the alignment of the local moments in presence 
of a magnetic field and the residual antiferromagnetic exchange. Magnetic 
transitions are known to occur in the half filled Hubbard model~\cite{krauth,zhang} 
in the insulating state. Of course, the problem is more involved in 
the metallic side due to the presence of an 
additional energy scale (the local Kondo temperature), associated  with the 
quenching of local spin fluctuations at low temperature. So, the question
raised above is much more subtle in the metallic phase of the Mott-Hubbard
systems. A quantitative study of the dynamics and transport properties of this 
model is not easy. Recently, much progress has been made with the advent 
of dynamical mean field theory~\cite{vollhard,pruschke,georges,
gebhard} which is exact in the limit of infinite 
spatial dimension. Within DMFT, a correlated model maps onto a single 
impurity model in a self-consistent conduction-electron bath. 
Methods like Numerical renormalization group (NRG)~\cite{pruschke1}, 
Local moment approach (LMA)~\cite{logan,logan1,parih}, Exact diagonalization 
(ED)~\cite{georges}, quantum Monte Carlo (QMC)~\cite{georges}, 
Iterative perturbation  theory (IPT)~\cite{kajueter,bulla,raja1} have been 
used to 
study the impurity problem within the DMFT framework. Mott-transition has been 
observed in the half filled Hubbard model using DMFT~\cite{krauth,zhang}. 

Prior to the advent of DMFT, there have been two major approaches for the 
strongly correlated Fermi liquid - the Stoner approach~\cite{anderson,beal,levin} and the 
Gutzwiller approach (GA)~\cite{vollhardt1,spalek,korbel}.
The former approach views the system as close to a ferromagnetic transition 
and the latter views it as a strongly correlated, nearly incompressible 
Fermi liquid close to a Mott localization. While the Stoner theory is 
essentially a high temperature approach, the Gutzwiller method starts from 
a Hubbard model description of the strongly correlated electrons and projects 
out the local double occupancies from the ground state in a variational calculation. The GA is incapable of describing the metallic phase of the half-filled
Hubbard model, and also fails to incorporate any dynamics or 
temperature dependence. However, it does have the merit of being analytically
tractable, and thus can yield qualitative insight, provided the results
have been benchmarked against other detailed calculations.
These two approaches have very different predictions for the response to 
an external magnetic field. The Stoner approach predicts a smooth variation 
of magnetization in presence of a magnetic field while the 
Gutzwiller approach gives a first order transition of the magnetization.  
Experimentally, an effort to distinguish between the two predictions,
based on studies in liquid $^{3}He$~\cite{wolf}, finds a smooth variation
of the magnetization as predicted in Stoner approach. Within DMFT,
calculations~\cite{krauth1} using exact diagonalization (ED) as the 
impurity solver were carried out to study the effect of magnetic field 
on the Hubbard model. A fictitious temperature scale was introduced 
through a frequency cutoff. The calculations were carried out at a fixed 
magnetization. A metamagnetic transition was obtained, and the $T-U$ `phase
diagram' was extended to the finite field axis, that demonstrated 
a similarity between the effects of temperature and field. A qualitative
agreement with the Guzwiller approach was also found. Later, an exchange
term was introduced~\cite{geor2} to incorporate the spin-spin correlation to 
combine features of Mott localization and almost ferromagnetic Stoner 
instability. Changes in dynamics with applied field, however, was not studied 
in these papers. 
A recent study of the same problem was carried out by Bauer~\cite{bauer}
using the numerical renormalization group method within DMFT. The focus 
in this study was on the metamagnetic transition, and the results
were qualitatively similar to the earlier work; the main result being
that the divergence of susceptibility at the metamagnetic transition is
not a consequence of effective mass divergence, but occurs through
the quasiparticle interaction term. The prediction for the coexistence
region in the $h-U$ plane was not made explicitly in the NRG study,
nevertheless, it is possible to deduce this, from the results presented.
There appears to be  a strong disagreement between the NRG and ED study
in this context.

In this paper, we study the effects of magnetic field on correlated
electron liquids within the framework of the half-filled Hubbard model,
using the iterated perturbation theory (IPT) as the impurity solver for the
DMFT.  The relative merits and demerits of the IPT are discussed
 in the next section. Since DMFT is known to treat the local
dynamics quite accurately, it is expected to reveal the local nature of the
competition between the spin fluctuation and the aligning field.  
We calculate the single particle dynamics, magnetization and other properties 
of the Hubbard model in the presence of an external magnetic 
field. In the infinite dimensional limit, the magnetic field appears only
as a Zeeman term and there is no explicit orbital contribution in the 
Hamiltonian as the lattice model is mapped onto a single-impurity model. 
There are no non-local terms, and the effects due to the non-interacting 
medium are included as a dynamical mean field in the hybridization. 
Our focus in this study is on the field-induced changes in quasiparticle 
weight and spectral
functions as the interaction strength is varied. Although
the DMFT+NRG study by Bauer~\cite{bauer} does examine dynamics, the detailed
behaviour of the crossover from the low field Fermi liquid behaviour
to the high-field non-Fermi liquid behaviour was not examined. Such
a crossover was proposed in the DMFT+ED work~\cite{krauth1}, but
was not supported by dynamics. Here we provide a comprehensive picture
of this crossover. In our study, we show the detailed behaviour of
the quasiparticle weights, that show an asymmetric cusp
like singularity in the effective mass as a function of field. We show 
that the effective mass increases monotonically as the metamagnetic 
transition is approached, and beyond the transition the effective mass 
decreases smoothly and monotonically for lower $U$, while for higher $U$, 
there is a sharp discontinuous decrease. The applied field gets renormalized
to an effective field through the polarization of the medium (within DMFT)
and via interaction effects.

We show that the lattice Kondo resonance at the Fermi level 
splits into two at large fields: the up and down components move away 
from the Fermi level 
and finally form a spin polarized band insulator. The shift of the band 
is not rigid due to the competition between the Kondo screening and the 
Zeeman effect. 
The hysteresis across the metamagnetic transition is used to find
the `phase diagram'. Our calculations agree quantitatively with the NRG
study of Bauer~\cite{bauer} and we regard this agreement as an important 
benchmark. The low field universal features of the spectral functions
are carefully examined and adiabatic continuity to the non-interacting
limit is demonstrated. We argue that the field-induced metamagnetic 
transition is from 
a correlated paramagnetic metallic liquid to a band insulator, and in 
this sense,
is very different from the interaction driven first order Mott transition,
which is from a correlated metallic liquid to a 
correlated Mott insulator. 
We compare our theoretical result with 
the experimental one on liquid $^{3}He$ and find good qualitative agreement.  

The paper is organized as follows. We begin in Sec.II with a brief 
description of the model. In Sec.III, we present our theoretical results 
and their analysis. We also compare our theory with the experimental results
on liquid $^{3}He$. Finally, we conclude 
in Sec.IV.

\section{Model and formalism}

The single band Hubbard model Hamiltonian describing correlated electrons 
in the presence of an external magnetic field 
(the orbital contribution is neglected) is given by: 
\begin{align}
\hat{H}=&-t\sum_{\langle i,j\rangle ,\sigma}(c^{\dagger}_{i\sigma}
c^{\phantom{\dagger}}_{j\sigma} + h.c)+U\sum_{i}n_{i\uparrow}n_{i\downarrow}\nnu \\ 
&+\sum_{i\sigma}(\ep_c-h\sigma)n_{i\sigma}
\label{eq:hubbard}
\end{align}
\noindent The first term describes the kinetic energy of the non-interacting
conduction ($c$) band due to nearest neighbor hopping $t$. The second term 
refers to the on-site repulsion $U$. The final term represents the orbital 
energy and the Zeeman splitting in an external magnetic field. For the 
particle-hole symmetric case (half-filled) case considered in this work,
the orbital energy is given by $\ep_c=-U/2$. In the limit
 of large dimensions, $D\rightarrow \infty$, the hopping needs to be scaled as 
$t\propto t_*/\sqrt{D}$.
 We choose, for convenience, a hypercubic lattice, for which 
the non-interacting density of states is an unbounded Gaussian 
($\rho_{0}(\epsilon)= \exp(-\epsilon^{2}/t^{2}_{*})/(\sqrt{\pi} t_{*})$). 
We set $t_{*}= 1$ in the following calculation. 

Within dynamical mean field theory (DMFT), which is exact in the limit of 
infinite dimensions, as is well known, the lattice model
may be mapped to an effective single-impurity Anderson model with a 
self-consistent hybridization. The major simplification that occurs
is that the self-energy and the vertex function become local and
momentum-independent. 
In the presence of an external global magnetic field, the 
local, retarded Green's function is given by, 
\begin{align}
G_{\sigma}(\omega,h)=H\left[\gamma_\sigma(\om;h)\right] 
\label{eq:gf}
\end{align}
where $\gamma_\sigma(\om;h)=\omega^+ - \ep_c + \sigma h - 
\Sigma_{\sigma}(\omega,h)$, $H[z]$ is the Hilbert trasform defined
as
\begin{displaymath}
H\left[z\right]=\int\frac{\rho_{0}(\ep)d\ep}{z-\ep}
\end{displaymath}
and
$\Sigma_{\sigma}(\omega;h)$ is the spin-dependent local self-energy.
The local Green's function may be used to define a hybridization function
$S(\om;h)$ as 
\begin{align}
G_{\sigma}(\omega,h)=\frac{1}{\gamma_\sigma(\om;h)-S(\om;h)}
\label{eq:gf2}
\end{align}
or, expressed another way,
\begin{equation}
S(\om)=\gamma_\sigma(\om;h) - \frac{1}{H\left[\gamma_\sigma(\om;h)\right]}
\end{equation}

\noindent
The hybridization function defines the host or the medium within which the
impurity is embedded, and the non-interacting host Green's function is 
then given by
\begin{align}
\cg_{\sigma}(\omega,h)=\frac{1}{\om^+ - \ep_c+\sigma h - S(\om;h)}
\label{eq:host}
\end{align}
which is equivalent to using the Dyson's equation
\begin{align}
\cg^{-1}_{\sigma}(\omega,h)=G^{-1}_{\sigma}(\omega,h)+\Sigma_{\sigma}(\omega,h)
\,.
\label{eq:gfs}
\end{align}
This host Green's function is then used to construct a new impurity
self energy, which if substituted in equation~\ref{eq:gf}, yields
a new lattice Green's function. Thus the solution within DMFT proceeds
most easily in an iterative manner using equations (2)-(6) until
self-consistency is achieved.

 Although easily stated,
constructing a new impurity self-energy for an arbitrary non-interacting host
has remained a major bottleneck in the DMFT scheme. Various 
impurity solvers have been developed, most important of which are
Quantum Monte Carlo, exact diagonalization, numerical renormalization group,
diagrammatic perturbation theory based approaches etc. We employ
the iterative perturbation theory (IPT) approach to solve the 
impurity problem within dynamical mean field theory (DMFT). The 
IPT is a very simple yet powerful diagrammatic perturbation theory
based approach. It has been benchmarked extensively against more
exact methods in the zero field half-filled Hubbard model problem.
The agreement, though qualitative, is excellent. The main advantage
of IPT is the ability it provides to capture real-frequency dynamics, 
and hence transport on all frequency, temperature scales and interaction
strengths.  For a more detailed comparison with other impurity solvers,
we refer the reader to a previous article by one of the authors
of the present work~\cite{raja1}. 

The zero temperature ($T=0$) self energy has the following form,
consisting of the static Hartree contribution and the dynamical part,
\begin{align}
\Sigma_{\sigma}(\omega,h)=U{\bar n}_{\bar{\sigma}}+
\Sigma_{\sigma}^{(2)}(\omega,h)
\label{eq:self}
\end{align} 
where ${\bar n}_{\sigma}=\int_{-\infty}^{0}
d\omega\, {\cal D}_\sigma(\om)$ and ${\cal D}_\sigma(\om)= -\frac{1}{\pi}
{\rm Im} {\cal G}_{\sigma}(\omega)$ is the spectral function of the host Green's
function. 
The local IPT self energy $\Sigma_{\sigma}^{(2)}(\omega,h)$ which satisfies the 
Luttinger theorem automatically at half-filling is given by (this zero field
algorithm has been discussed recently in detail~\cite{raja2}),
\begin{align}
\Sigma_{\sigma}^{(2)}(\omega,h)=U^{2}\int_{-\infty}^{\infty}
d\omega^\prime {\cal D}_{\bar{\sigma}}(-\om;h)\times \nnu \\
\left[\chi_1^{\sigma{\bar{\sigma}}}(\om+\om^\prime)f(\om^\prime)+
\chi_2^{\sigma{\bar{\sigma}}}(\om+\om^\prime)f(-\om^\prime)\right]
\label{eq:self1}
\end{align}
where 
\begin{align}
\chi_1^{\sigma\bar{\sigma}}(\omega)=\int^\infty_{-\infty}{d\omega^\prime}
{\cal D}_\sigma(\om^\prime){\cal D}_{\bar{\sigma}}(\om+\om^{\prime})
f(\om^\prime)f(-\om-\om^\prime) \nnu \\
\chi_2^{\sigma\bar{\sigma}}(\omega)=\int^\infty_{-\infty}{d\omega^\prime}
{\cal D}_\sigma(\om^\prime){\cal D}_{\bar{\sigma}}(\om+\om^{\prime})
f(-\om^\prime)f(\om+\om^\prime) 
\label{eq:pola}
\end{align} 
The magnetization can be calculated from  
\begin{align}
m=\langle n_{\uparrow}\rangle-\langle n_{\downarrow}\rangle 
\label{eq:mag}
\end{align}
where $\langle n_{\sigma}\rangle =\int_{-\infty}^{0}
d\omega\, (-\frac{1}{\pi}){\rm Im}G_\sigma(\om)$.

\section{Results and discussions}
In this section, we will discuss the results of our calculations so far on 
the effect of magnetic field on Hubbard model at half filling. Before discussing 
the finite field results, we will briefly review the well-known zero field results 
for the Hubbard model. At zero field, the system goes from a metal to an insulator 
at $U>U_{c_{2}}$. Below $U_{c_{1}} (< U_{c_{2}})$, the system is metal and shows 
Fermi liquid behaviour. There is a coexisting region between $U_{c_{1}}$ 
and $U_{c_{2}}$. The values 
of $U_{c_{1}}$ and $U_{c_{2}}$ for the hypercubic lattice are 3.67$t_{*}$ and 
4.78$t_{*}$~\cite{krauth,zhang}. We are interested in the region  
below $U_{c_{2}}$. At zero field, the spectra shows 
an universal scaling form in terms of $\omega_{L}$ (where $\omega_{L}=Zt_{*}$
and Z is the quasiparticle weight). 

We will begin with a discussion of the symmetry properties of the Green's
function and the associated self energies. The following
symmetry holds in the half-filled case:
\begin{align}
G_\sigma(\om;h)=-\left[G_{\bar{\sigma}}(-\om;h)\right]^*  \\
\Sigma\sigma(\om;h)=-\left[\Sigma_{\bar{\sigma}}(-\om;h)\right]^* 
\label{eq:symm}
\end{align}
To see this, imagine carrying out the self-consistent DMFT iterations beginning
with a symmetric hybridization function, as appropriate for the zero-field
case. Naturally, the resulting host Green's function (equation~\ref{eq:host})will satisfy the above symmetries.
The self-energies calculated through the IPT ansatz 
(equations~\ref{eq:self1}-\ref{eq:pola})
would then also satisfy the same symmetries, as then would the full interacting
Green's function (equation ~\ref{eq:gf2}). One consequence of such a symmetry is that the quasiparticle
weights $Z_\sigma$ would be spin-independent.

Rewriting equation~\ref{eq:self} as 
$
\Sigma_\sigma(\om;h)=\frac{U}{2} ({\bar{n}} -\sigma {\bar{m}}) + \Sigma^{(2)}_\sigma
(\om;h)
$
where ${\bar{m}}= {\bar{n}}_\upa - {\bar{n}}_\dna$ is a 
fictitious local moment (of the host/medium);
and using $\ep_c=-U/2$ and ${\bar{n}}=1$, we 
transform Eq.~\ref{eq:gf} as, 
\begin{align}
G_\sigma(\om;h)=\int^\infty_{-\infty} \frac{\rho_0(\ep)\,d\ep} {\om^+ -\ep
+ \sigma\left(h+\frac{U}{2}{\bar{m}}\right) - \Sigma^{(2)}_\sigma(\om;h)}
\label{eq:gf3}
\end{align}
We would like to mention a technical point here. We found that the IPT equations are ill-behaved for a fixed field. In other words, if we wish to find the
value of the magnetization self-consistently through the iterative procedure,
keeping the field fixed, we run into problems of convergence. An easy way to 
avoid this instability is to work with a fixed value of the composite number 
$h_m\equiv h+U{\bar{m}}/2$. The equations converge easily and a unique solution is found. The host Green's functions can be then used to find ${\bar{m}}$, and the
field value is obtained as $h=h_m - U{\bar{m}}/2$.  A similar strategy was 
employed by 
Laloux et al~\cite{krauth1} in their ED+DMFT approach, where they had to use a fixed
magnetization to achieve convergence.
The final value of the actual magnetization is found through 
equation~\ref{eq:mag}.
\begin{figure}[t]
\centerline{\hbox{
\epsfig{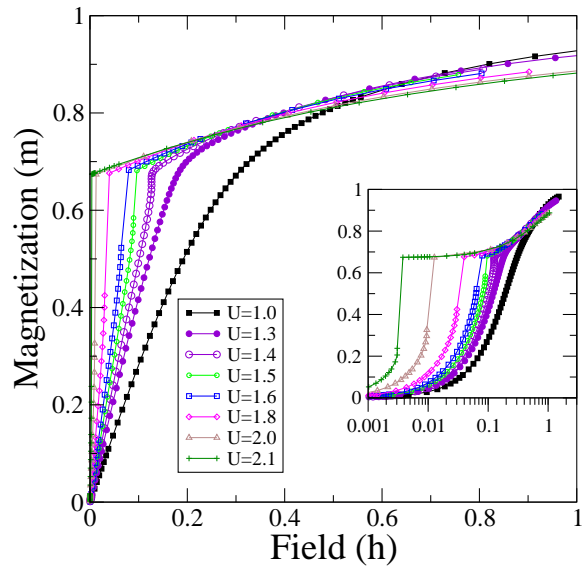}}}
\caption{(Color online) Magnetization as a function of
field for various $U/t_*$ values. 
The magnetization smoothly approaches
the saturation value of $1$ for $U<1.4t_*$, while for $U>1.4t_*$,
a metamagnetic first order transition is seen. The magnetization 
jumps at a finite value of the field, which falls sharply with
increasing $U/t_*$ (see figure~\ref{fig:hcU}), and beyond $U/t_*=2.2$ becomes 
unresolvably small
within the present approach. The inset shows the same data as the main panel
on a logarithmic field axis, so that the sharp decrease in the $h_c$ value 
is  seen clearly. }
\label{fig:mag_h}
\end{figure}

\begin{figure}[h]
\centerline{\hbox{
\epsfig{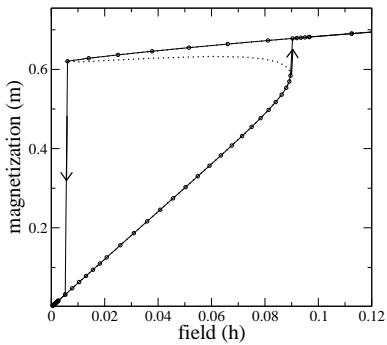}}}
\caption{Hysteresis is seen across the metamagnetic transition. 
We illustrate this for $U=1.5t_*$ here. The dotted line is the actual
obtained line when the parameter $h_m$ is continuously increased. The region 
where the susceptibility, $\chi=\partial m/\partial h$ becomes negative 
is excluded to obtain the
hysteresis.
}
\label{fig:hyst}
\end{figure}

\begin{figure}[ht]
\centerline{\hbox{
\epsfig{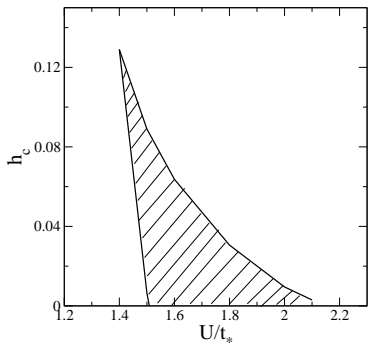}}}
\caption{The coexistence  region (shaded) as obtained within IPT for the
hypercubic lattice. At large
$U (\gtrsim 2.2t_*)$, the system is susceptible to being polarized for an infinitesimal
field.
}
\label{fig:hcU}
\end{figure}

Our results for magnetization are summarized in figure~\ref{fig:mag_h}.
The magnetization ($m$) is shown as a function of field ($h$) for increasing
values of the interaction strength $U/t_*$. 
The magnetization smoothly approaches
the saturation value of $1$ for $U<1.4t_*$, while for $U>1.4t_*$,
a metamagnetic first order transition is seen. The magnetization 
jumps at a finite value of the field $h_c$, which falls sharply with
increasing $U/t_*$ (figure \ref{fig:hcU}), and beyond $U/t_*=2.2$ becomes 
unresolvably small
within the present approach. The jump size increases from 0 at a 
$U_{hc}=1.4t_*$ to a large value ($\sim 0.67$) at $U=2.1t_*$. The value of
$U_{hc}=1.4t_*$ does agree quantitatively with that obtained by Bauer~\cite{bauer}
through NRG+DMFT. In the latter, a semi-circular density of states has been 
chosen to represent the bare conduction band, $\rho_0^{\rm sem} =
2\sqrt{D^2-\ep^2}/\pi D^2$ with $D=2t_*$. The
value of $U_{hc}/t_*$ was found to be 2.64 in the NRG study~\cite{bauer}, 
which corresponds
to a $U_{hc}/D\simeq 1.32$, and thus the agreement with the IPT value ($1.4$) for the
same provides our results an important benchmark. ($1.4$)  
This metamagnetic transition has been observed and studied previously within
the GA and DMFT approaches. The low field state is a paramagnetic state,
while the high field state beyond the transition is a polarized 
quasi-ferromagnetic state. 
As with the zero field case, the metamagnetic
transition being first order, would be accompanied by a hysteresis, as the 
field is swept through a closed cycle across the transition. This is shown 
in figure~\ref{fig:hyst} for $U=1.5t_*$. 
Thus, there exists, for
a range of $U$ values, a region in the $h_c-U$ plane that defines the
`coexistence' region, i.e the region, where the paramagnetic metallic
and the quasi-ferromagnetic insulating solutions coexist. This region 
may be identified through the bounds
of the hysteresis curve for each $U$. In figure~\ref{fig:hcU}, the shaded 
region represents the coexistence region, and the solid lines are the 
equivalents of $U_{c1}$ and $U_{c2}$ for the finite field case.
Indeed, Bauer~\cite{bauer} reports that
the metallic solutions were impossible to get for 
$U/t_*\gtrsim 4.5$, which corresponds
to $U/D\simeq 2.2$, which is the right boundary of the coexistence region 
in figure~\ref{fig:hcU}, and 
thus the coexistence region obtained within IPT for the hypercubic lattice 
agrees excellently with that of the NRG study. 
Since the quasi ferromagnetic high field insulator is not however a Mott 
insulator, the term 'equivalent' above is not strictly precise. In fact,
this two-phase region at finite field
is not an ``extension" of the $U_{c1}-U_{c2}$ region in the $h=0$ plane onto the
finite $h$ plane. They do not merge onto each other and represent very
different spinodal regions across different states.

\begin{figure}[t]
\centerline{\hbox{
\epsfig{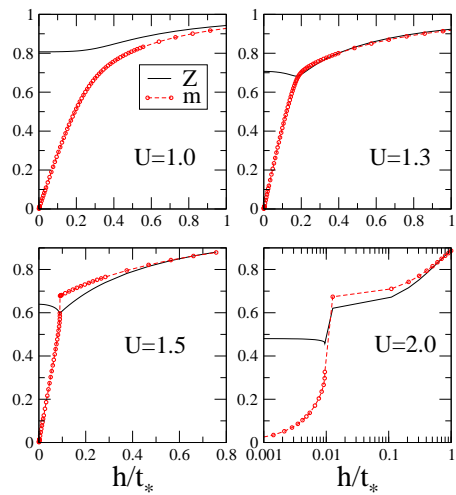}}}
\caption{The dependence of quasiparticle weight (inverse effective mass) is
illustrated here for four $U$ values. The cusp singularity is seen
to develop for intermediate and high $U$, and at precisely the metamagnetic
transition. 
}
\label{fig:ZM_U}
\end{figure}

We shift our focus to the quasiparticle weights or inverse effective mass.
In figure~\ref{fig:ZM_U}, we show the $Z$ as a function of field for 
various $U$ values, ranging from low $U$ at which metamagnetism is weak or
absent,
to a relatively high $U$, i.e the strongly metamagnetic region, 
where the system is susceptible to being
polarized even for an infinitesimal field. 

At low $U=1.0$, the magnetization increases smoothly with increasing field,
and the effective mass decreases monotonically, approaching the bare mass 
at high fields. The behaviour of $m$ and $Z$ changes qualitatively,
for $U=1.3t_*$; the effective mass displays a cusp, while the magnetization
changes slope significantly. The cusp sharpens for $U=1.5t_*$, while 
$Z$ suffers a discontinuous increase at the metamagnetic transition.
At even higher $U=2.0$, both the magnetization and effective mass display 
discontinuous changes at the transition. Thus the field-dependence of 
magnetization and effective mass is very sensitive to the specific $U/t_*$
values under consideration.

The main advantage with the IPT approach used
here, is that we can study the real frequency dynamics in detail. 
We now proceed to elucidate the field dependence of the dynamics before
and across the metamagnetic transition.
We begin by exploring the universal scaling properties by carrying out a
low frequency Fermi liquid analysis.
In the metallic phase of interest, the real part of self energy may be
expanded about the Fermi level to  first order in $\omega$,
\begin{align}
{\rm Re}\Sigma^{(2)}_{\sigma}(\omega,h)={\rm Re}\Sigma^{(2)}_{\sigma}(0,h)+
(1-\frac{1}{Z_{\sigma}(h)})\omega
\label{eq:fermi}
\end{align}
where $Z_{\sigma}(h)=[1-\partial\Sigma^{R}_{\sigma}(\omega;h)
/\partial\om]^{-1}$.  For the half-filled case considered in this paper,
$Z_\upa=Z_\dna$ (as argued before), hence we drop the spin subscript.
The imaginary part of the self energy remains Fermi liquid like
($\sim {\cal O}(\om^2)$) because of the structure of the IPT equations
(equations~\ref{eq:self1}-~\ref{eq:pola}). A finite imaginary part can
be acquired by the self energy only at finite temperatures in this approach.
Substituting the above in equation~\ref{eq:gf3}, the spectral function 
$D_{\sigma} = -{\rm Im}G_\sigma/\pi$, is obtained as 
\begin{align}
D_{\sigma}(\omega,h){\stackrel{\tilde{\om}\lesssim 1}{\longrightarrow}}\rho_{0}(\tilde{\om}+\sigma h^{eff})
\label{eq:nisptra}
\end{align}  
where $\tilde{\om}=\om/\om_{L}$ (with $\om_{L}=Z(h)t_{*}$) 
and the spin-independent (because of the symmetries, equations~\ref{eq:symm}) 
effective field is given by
\begin{align}
h^{eff}=h + {\frac{U}{2}}{\bar{m}} -\sigma {\rm Re}\Sigma^{(2)}_{\sigma}(0,h).
\label{eq:heff}
\end{align}
The above Eq.~\ref{eq:nisptra} is the renormalized non-interacting limit
of the full spectral function.
Given the above form of the spectral function, it is easy to see
that adiabatic continuity to the non-interacting limit is achieved
provided the real part of the self energy may be expanded as in
equation~\ref{eq:fermi}. Using equation~\ref{eq:nisptra}, we can infer 
that the spin-summed spectra would also be a universal function of 
$\tilde{\om}$ for a fixed $h_{eff}$. Whether the full field-dependent 
spectra satisfy adiabatic continuity and scaling may be tested through
equation~\ref{eq:nisptra}.

\begin{figure}
\centerline{\hbox{
\epsfig{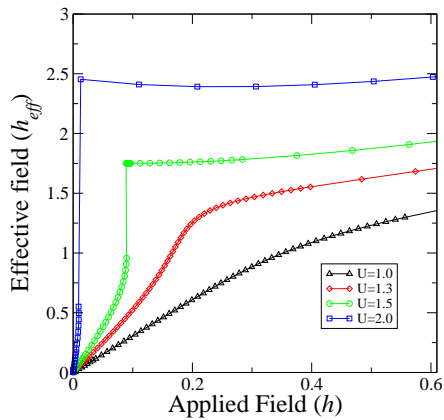}}}
\caption{(Color online) The effective field is plotted against applied
field ($h$) for the $U$ values chosen in figure~\ref{fig:ZM_U}.}
\label{fig:eff_h}
\end{figure}

The dependence of effective field ($h_{eff}$) on the applied field 
is shown in figure ~\ref{fig:eff_h} for various $U$ values.
The metamagnetic transition is reflected in the effective field as well.
A careful examination of the contribution to the effective field indicates that the discontinuity is present not only in the real part of $\Sigma^{(2)}(0;h)$,
but also in the host magnetization $\bar{m}$.

\begin{figure}
\centerline{\hbox{
\epsfig{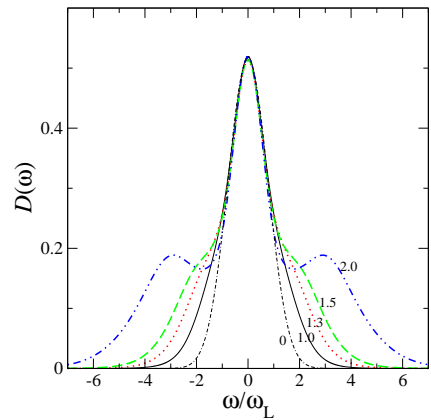}}}
\caption{(Color online) Adiabatic continuity and scaling: 
The full spectral function when plotted as a function of $\om/\om_L=
\om/(Z(h)t_*)$ for various $U$ (shown next to the curves) collapse in the
 neighbourhood of the Fermi
 level. The universal form is seen to be identical to the non-interacting
spectral function for the same parameters (see text for discussion).
} 
\label{fig:univ}
\end{figure}

The spectra for various $U/t_*$ plotted as a function of 
$\om/\om_L=\om/(Z(h)t_*)$ collapse in the neighbourhood of the Fermi level 
as shown in figure~\ref{fig:univ}. Also shown in the same graph is the 
non-interacting density of states (dos) for a field of $h=0.3t_*$ and $Z=1$. 
All the $U>0$ spectra are seen to be identical to that of the 
non-interacting case, near the Fermi level,
thus exhibiting adiabatic continuity and universal scaling as a function
of $\om/\om_L$, and validating equation~\ref{eq:nisptra}. This demonstrates
Fermi liquid behaviour for all $h$ lower than the $h_c$ for a given
$U/t_*$.

\begin{figure}[t]
\centerline{\hbox{
\epsfig{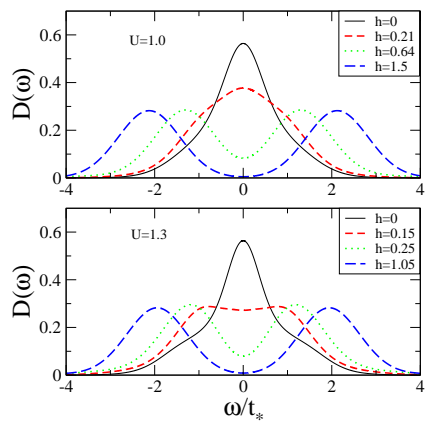} }}
\caption{(Color online) The upper (lower) panel represents the spectral 
function for $U=1.0$ ($U=1.3$) and various field values plotted against
the bare frequency, $\om/t_*$.
}
\label{fig:spec1}
\end{figure}

We now move on to the field dependence of the spectral functions for a 
given interaction strength. In the absence of clear metamagnetism ($U=1.0$,
the spectral function evolution with field is very similar to a non-interacting
case, as expected. The density of states at the Fermi level continues 
to decrease with increasing field, until the up and down spin bands
move so far from the Fermi level, that a band insulator is obtained. This is 
shown in the upper panel of figure~\ref{fig:spec1}. The lower panel of the same 
figure represents $U=1.3t_*$, which is close to the onset of metamagnetism.
Again, the behaviour is similar, albeit the decrease of the density of states
at the Fermi level with increasing field is more pronounced. 
We add here that the IPT results for the field-dependence of the spectral
functions are not in agreement with the NRG+DMFT results of 
Bauer et.\ al~\cite{bauer}. In the latter, the spectral function seems 
to be pinned at the Fermi level until the metamagnetic transition,
whence there is a sudden drop in the density of states. We do not understand
this difference with NRG completely, but speculate that this disagreement
could be due to the non-conserving nature of IPT.

\begin{figure}[h]
\centerline{\hbox{
\epsfig{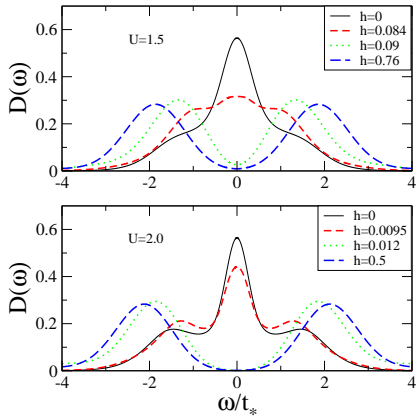} }}
\caption{(Color online) The upper (lower) panel represents the spectral 
function for $U=1.5$ ($U=2.0$) and various field values plotted against
the bare frequency, $\om/t_*$.}
\label{fig:spec2}
\end{figure}
The evolution of the spectral function  for $U=1.5$ and $U=2.0$ 
(figure~\ref{fig:spec2}) with field is quite dramatic, reflecting the
metamagnetic transition. The spin bands shift rapidly with field,
and for a critical field at which the transition happens, the shift
from a metallic to a quasi-ferromagnetic insulator takes place.
Though the insulating state is reminiscent of a Mott insulator, it is
actually far from being one. The two `Hubbard bands' are just the 
two spin bands shifted away from the Fermi level, and broadened to an extent
that they just resemble non-interacting Gaussian density of states. In fact,
the large-field spectral function may be deduced from the large field
asymptote of equation~\ref{eq:gf3}. At large field $h\sim t_*$, the effective 
field becomes
comparable to or larger than the largest scale in the problem, i.e $U$ 
in strong coupling. This is because, for $h\simeq t_*$, the magnetization
saturates, ${\bar m}\rightarrow 1$, hence the effective field becomes
at least $h+U/2$. The real part of the second-order self energy contributes
$1/U$ at frequencies $\om\simeq U$, because $\Sigma^{(2)}(\om)\sim 1/\omega$
at large frequencies. Thus from equation~\ref{eq:gf3} the large field
spectra must be simply given by
$\left[\rho_0(\om+h_{eff}) + \rho_0(\om-h_{eff})\right]/2$. This is just
a sum of two Gaussians each centred around $\pm h_{eff}$ respectively.
Indeed, from figure~\ref{fig:eff_h}, we see that for $U=2.0$, the value of
the effective field is $\sim 2.5$ for a field of $h=0.5$; and the lower
panel of figure~\ref{fig:spec2} shows that the two `Hubbard bands' are centred
around $\sim \pm 2.5$.
This indicates, that at high field, all correlation effects are lost,
and the `Hubbard bands' are nothing but the bare density of states
shifted away from the Fermi level, and thus the insulator at high fields 
is just a band insulator.

\begin{figure}[h]
\centerline{\hbox{
\epsfig{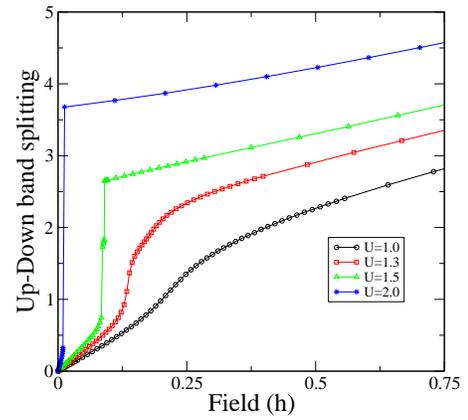} }}
\caption{(Color online) The splitting between the up and down spin bands
is plotted against field for various $U$ values.
}
\label{fig:split}
\end{figure}

The split between the up and down spin bands may be quantified, and is
shown in figure~\ref{fig:split} for various $U$ values. The field induced
shift of the spin bands is seen to be highly non-linear and indeed even
discontinuous for $U=1.5$ and $U=2.0$, naturally at the metamagnetic
transition. 
In the non-interacting case this split is a rigid Zeeman shift. But 
in the interacting case the shift is not rigid at all.  In the periodic
Anderson model (appropriate for heavy fermion
 systems), the effect of magnetic field in the Kondo lattice regime
is similar (without metamagnetism), and it has been argued that the 
non-rigid shift is due
to the competition between Kondo screening of the local moments
(whose effect is to quench the local moments) and the Zeeman coupling
(whose effect is to polarize the local moments). In the present
case, since the lattice Hubbard model is mapped onto the single
impurity Anderson model within DMFT, similar arguments must apply. 
And it is possibly this competition here, that is ultimately responsible
for the first order metamagnetic transition. We see that the field
induced transition is very different from the temperature induced transition,
or the pure interaction induced Mott transition at zero field. In the latter
especially, as one approaches $U_{c2}$ from the metallic side, the
Fermi liquid scale $Zt_*$ vanishes continuously, and upon crossing into
the Mott insulator side, a large and finite gap appears discontinuously
in the density of states. At $U_{c1}$ the reverse happens, i.e the
gap in the Mott insulator vanishes continuously on decreasing $U$,
and on crossing $U_{c1}$, a Fermi liquid with a finite quasiparticle weight 
$Z$ is found. At finite field, however, neither the Fermi liquid
scale, nor the gap, vanishes at the metamagnetic transition. 

\begin{figure}
\centerline{\hbox{
\epsfig{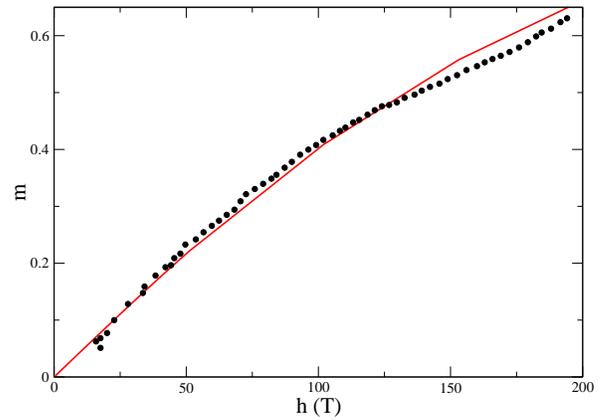} }}
\caption{(Color online) Comparison of the experimental magnetization in liquid 
$^{3}He$ (filled circle) with theory (solid line).
}
\label{fig:expt}
\end{figure}
Finally we would like to compare our theory 
with experiments. The experimentally measured magnetization~\cite{wolf}
 as a function of effective field at temperatures $T\sim 70-90 mK$ 
is shown in figure~\ref{fig:expt} as 
filled circles. The authors of the experimental paper concluded that
the first order metamagnetic transition scenario proposed by 
Vollhardt~\cite{vollhardt1} cannot
be applicable to the case of liquid $^3$He, since metamagnetism
is absent in the experimental data, although a slight metamagnetic
kink can be distinguished at a field of about 180T. We
argue that
the applicability of the half-filled Hubbard model cannot be ruled
out yet, since as our figure ~\ref{fig:mag_h} shows, the magnetization
can be a smooth function of the field for low to intermediate interaction
strengths. According to Vollhardt~\cite{vollhardt1}, $U$ is typically 
about 15K for $^3$He, while
the parameter $U/E_F \sim 0.8$, where $E_F$ is the Fermi energy. 
So this does seem to be an intermediate coupling
scenario. Indeed, if we superimpose the calculated magnetization for
$U=1.0t_*$ onto the experimental data (after a multiplicative scaling
of the field axis), we see surprisingly good agreement. The Gutzwiller 
approach yields a discontinuous metamagnetic transition for rather
weak interactions, while we find that metamagnetism does not develop until
a reasonably strong interaction strength $U\gtrsim 1.3t_*$ within the 
IPT+DMFT approach employed here.  

\section{Conclusions}
In this paper, we have studied the effect of magnetic field on correlated
electron liquids within the framework of the 
half filled Hubbard model using
a dynamical mean field theory. We compute the  
the single particle dynamics, magnetization and other zero 
temperature properties. For intermediate $U$ values, we find a first order
itinerant metamagnetic transition, in agreement with earlier theoretical studies.
An important benchmark for our study was the agreement with the 
NRG+DMFT calculations~\cite{bauer} regarding the onset of metamagnetism and the 
coexistence region in the $h-U$ plane. We conjecture that the first order 
metamagnetic
transition is a result of the competition between Kondo screening and
Zeeman coupling. The metamagnetic transition is found to be very different 
in nature from the zero-field interaction driven Mott transition. 
Beyond $U=2.2t_*$, we find that even an (numerically) 
infinitesimal field is sufficient to push the system from a zero field 
paramagnetic state to a polarized quasi-ferromagnetic state with a substantial
magnetization. Finite temperature transport and dynamics studies would
be instrumental in furthering an understanding of this transition.

\section{Acknowledgments}
D.P would like to thank Prof. T. V. Ramakrishnan for financial support through
his grant. We would like to acknowledge enlightening discussions with
Prof.\ H.\R.\ Krishnamurthy.

\end{document}